# THE PARTICLE ENIGMA

Gerald E. Marsh
Argonne National Laboratory (Ret)
gemarsh@uchicago.edu

**ABSTRACT**

The idea that particles are the basic constituents of all matter dates back to ancient times and formed the basis of physical thought well into modern times. The debate about whether light was a wave or a stream of particles also lasted until relatively recently. It was the advent of de Broglie's work and its implications that revolutionized the concept of an elementary particle—but unfortunately did not banish the idea of a point particle despite its difficulties in both classical and quantum physics. Some of these problems are discussed in this essay, which covers chiral oscillations, Penrose's "zigzag" picture of particles satisfying the Dirac equation, and some ideas derived from string theory.

What is a particle? We all know that the concept of a particle comes from Democritus' idea of atoms. His conception, and what today we would call Brownian motion, was related by Lucretius to the origin of all motion in his poem *On the Nature of Things* (50 B.C.E.):

> *Whence Nature all creates, and multiplies*
> *And fosters all, and whither she resolves*
> *Each in the end when each is overthrown.*
> *This ultimate stock we have devised to name*
> *Procreant atoms, matter, seeds of things,*
> *Or primal bodies, as primal to the world.*
> • • •
> *For thou wilt mark here many a speck, impelled*
> *By viewless blows, to change its little course,*
> *And beaten backwards to return again,*
> *Hither and thither in all directions round.*
> *Lo, all their shifting movement is of old,*
> *From the primeval atoms; for the same*
> *Primordial seeds of things first move of self,*
> *And then those bodies built of unions small*
> *And nearest, as it were, unto the powers*
> *Of the primeval atoms, are stirred up*
> *By impulse of those atoms' unseen blows,*
> *And these thereafter goad the next in size;*
> *Thus motion ascends from the primevals on,*
> *And stage by stage emerges to our sense,*
> *Until those objects also move which we*
> *Can mark in sunbeams, though it not appears*
> *What blows do urge them.*

With a little license, Lucretius' "*Procreant atoms, matter, seeds of things, Or primal bodies*" formed the basis of physical thought until quite late into modern times. In the ancient world, however, while it was accepted there might be different kind of atoms, the number of types was small and sometimes related to geometrical shapes. The advent of modern chemistry and spectroscopy in the 19th century began the formation of the current understanding of the nature of atoms.

Today, it is believed that the elementary building blocks of matter are leptons and quarks, all of which are called fermions and obey the Dirac equation for a particle of spin of ½. In addition, there is electromagnetic radiation carrying a spin of 1. Lucretius' understanding of atoms has been carried over into the modern conception of "particle" in the sense that the basic fermions are thought to be "structureless" or "point" particles. This can be seen in the attempts to construct "classical" models for the electron. Examples are the de Broglie-Bohm interpretation of quantum mechanics[1] and the work of David Hestenes.[2] But retaining the idea of a massive charged point particle requires that both mass and charge be renormalized, a process that has never rested comfortably with many physicists.

The greatest challenge to the ancient idea of a particle came from the work of de Broglie, who introduced in 1924 the idea that each particle had associated with it an internal clock of frequency $m_0c^2/h$. From this idea he found his famous relation showing particles of matter were associated with a wave.[3] He did not believe a particle like the electron was a point particle, but rather that the energy of an electron was spread out over all space with a strong concentration in a very small region: "L'électron est pour nous le type du morceau isolé d'énergie, celui que nous croyons, peut-être à tort, le mieux connaître; or, d'après les conceptions reçues, l'énergie de l'électron est répandue dans tout l'espace avec une très forte condensation dans une région de très petites dimensions dont les propriétés nous sont d'ailleurs fort mal connues."[4]

**1. The de Broglie Relation: Theory and Experiment**

De Broglie, in his 1929 Nobel lecture used the following argument:

$$p = \gamma m_0 v = \gamma m_0 c^2 \frac{v}{c^2} = E \frac{v}{c^2}.$$

Identifying the energy of the *massive particle* with $E = h\nu$ gives

$$p = \frac{h\nu}{c^2/v}.$$

De Broglie then assumed that $c^2/v$ corresponds to a phase velocity via $vV = c^2$, so that

$$p = \frac{h\nu}{V} = \frac{h}{V/\nu}.$$

Using $V = \nu\lambda$, he obtains his relation $\lambda p = h$.

Note that by assuming that $c^2/v$ corresponds to a phase velocity de Broglie is introducing waves having neighboring frequencies so that he can define both phase and group velocities. The phase velocity so introduced is, in Max Born's words, "a purely artificial conception, inasmuch as it cannot be determined experimentally."[5]

The existence of de Broglie's internal clock has recently been directly subject to experiment. The experimental approach used is known as "electron channeling", a phenomenon observed in silicon crystals.[6,7] In the experiments, a stream of electrons is aligned along a major axis of a thin single crystal corresponding to a row of atoms. The transmission probability along this axis, compared to neighboring angles with respect to the axis, is reduced except for a sharp peak in the direction of the axis. This peak or resonance occurs at a momentum of ~80MeV/$c$ as can be seen in Fig.(1.1)[8]. The electrons responsible for this resonance move along the cylindrical potential of one row of atoms in precessing orbits known as Sommerfeld rosettes—they are captured in a bound state of the row's potential. The position of this resonance corresponds to the de Broglie frequency.

While the experimental data show a resonance at a momentum of ~80MeV/$c$, modeling scattering calculations predict a resonance at ~161MeV/$c$. The discrepancy between the model results and the experimental results have not been completely resolved as of this writing.

What is clear, however, is that the electron momentum of ~80MeV/$c$ corresponds to a momentum of $2.7 \times 10^{-22}$ $kg\ m/s$, and using the relativistic expression for the energy, $E^2 = p^2c^2 + m_0^2c^4$, this corresponds to an energy of $1.2 \times 10^{-11} j$. This allows the calculation of the de Broglie wavelength to be $\lambda_{deB} = 1.6 \times 10^{-14}$ $m$, corresponding to a frequency of $\nu_{deB} = 1.8 \times 10^{22}$. Note that in the rest frame of the moving electron, these values must be compensated for the relativistic motion using $\gamma = 156.5$ giving $\lambda_{deB} = 2.5 \times 10^{-12}$ $m$ and $\nu_{deB} = 1.15 \times 10^{20}$, the usual values for $\lambda_{deB}$ and $\nu_{deB}$ in the rest frame. The atomic spacing in the crystal lattice is $d = 3.8 \times 10^{-10}$ $m$, significantly larger than the de Broglie wavelength.

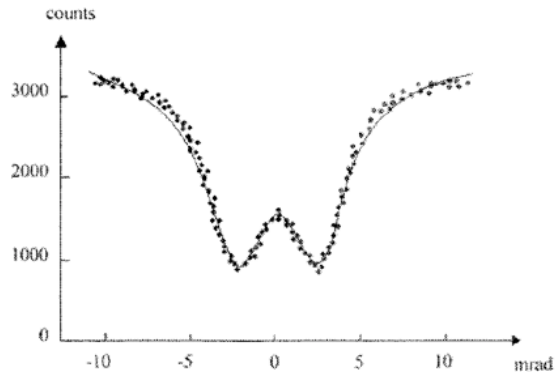

Figure (1.1). Experimental data from electrons of momentum 80 MeV/$c$ aligned along the <110> direction of a 1 $\mu$m thick silicon crystal. The figure shows counting rate vs. crystal tilt angle. The full curve is the product of a phenomenological calculation. [Adapted from P. Catillon, et al. *Found Phys* **38**, 659-664 (2008).

Appendix 1 has a discussion of the theory behind the channeling resonance phenomena given by Bauer.[9]

## 2. The Zig-Zag Picture of the Electron

Penrose has introduced another conception of particles satisfying the Dirac equation and specifically applied it to the electron—he calls it "The zigzag picture of the electron".[10] He continues, however, to view fermions like the electron as point particles.

Penrose's idea is related to the concept of zitterbewegung, which is generally associated with a point particle showing that the minimum effective size of the particle is its Compton radius. While the conventional phenomenon of zitterbewegung has been covered in many textbooks, it may be useful to have a presentation on it here. Appendix 2 covers this as well as the equivalent for electromagnetic radiation, a phenomenon that is not so well known.

The electroweak interactions inform us that the fundamental constituents of matter are leptons and quarks. These irreducible building blocks of spin ½ fermions are fields that transform under the left and right handed (or spinor) representations of the Lorentz group designated by (½, 0) or (0, ½ ) respectively. Thus a Dirac fermion field combines two equal mass 2-component fields into the group $(1/2, 0) \oplus (0, 1/2)$, which is a reducible representation of the Lorentz group.[11]

The Dirac 4-spinor can then be represented as a pair of 2-spinors (call them $\phi_L$ and $\phi_R$) and the Dirac equation becomes an equation coupling these two spinors with the coupling constant being related to the mass of the particle. These equations can be written

$$(p_0 + \boldsymbol{\sigma} \cdot \boldsymbol{p}) \phi_L(\boldsymbol{p}) = m \phi_R(\boldsymbol{p}),$$
$$(p_0 - \boldsymbol{\sigma} \cdot \boldsymbol{p}) \phi_R(\boldsymbol{p}) = m \phi_L(\boldsymbol{p}),$$

(2.1)

where the 4-vector $p^\mu$ = (E, $\boldsymbol{p}$), $p_\mu$ = (E, –$\boldsymbol{p}$), and the signature is –2. The quantity $\boldsymbol{\sigma} \cdot \boldsymbol{p} / |\boldsymbol{p}| = \boldsymbol{\sigma} \cdot \hat{\boldsymbol{p}}$ is the component of spin in the direction of the momentum and is the helicity. When the mass vanishes, helicity is the same as chirality where one also speaks of left and right chirality, but this can be quite misleading especially when the mass is not zero. In that case chirality is a purely quantum mechanical quantity related to the weak interactions, which do not exhibit mirror reflection symmetry.

Penrose calls these 2-spinors "zig" and "zag" particles, which he assumes to be massless. He furthermore assumes that these particles are continuously converting themselves into each other at a frequency related to the zitterbewegung frequency, which he observes is "essentially the de Broglie frequency". There is a problem with this in that the de Broglie and zitterbewegung frequencies differ by a factor of two. There is also the additional problem that chiral oscillations vanish for massless particles (see discussion just above Eq. (2.5) below). Penrose is obviously aware of this when he states that for these massless particles, "Each is the source for the other, with the rest-mass as coupling constant".

The left-handed massless zig particle is defined by Penrose as that part of the Dirac field projected out by the operator ½(1– $\gamma^5$), having helicity –1/2, and the right-handed zag particle as that projected by ½(1+ $\gamma^5$) having helicity ½. In the rest frame of the electron the zigzag oscillation is at the speed of light with the direction of spin remaining constant. It is important to keep in mind that for mass-zero particles helicity is the same as chirality and these projection operators are really the chirality operators.

The zigzag oscillation is a chiral oscillation. The way to see this is by decomposing the Dirac wave function as

$$\psi(x) = \psi_L(x) + \psi_R(x) \equiv \tfrac{1}{2}(1-\gamma^5)\psi(x) + \tfrac{1}{2}(1+\gamma^5)\psi(x).$$

(2.2)

The symmetries can be found by use of the Lagrangian

$$\mathcal{L} = \overline{\psi}(i\partial\!\!\!/ - m)\psi = \overline{\psi}_L i\partial\!\!\!/\psi_L + \overline{\psi}_R i\partial\!\!\!/\psi_R - m(\overline{\psi}_L\psi_R + \overline{\psi}_R\psi_L).$$

(2.3)

Here, $\overline{\psi}$ is the Dirac adjoint defined as $\overline{\psi} = \psi^\dagger \beta = \psi^\dagger \gamma^0$, and $\partial\!\!\!/ = \gamma^\mu \partial_\mu$. Note that the kinetic energy connects $L$ to $L$ and $R$ to $R$, while the mass terms connect $L$ to $R$ and $R$ to $L$.

For $m \neq 0$, $\psi \to e^{i\theta}\psi$ leaves this Lagrangian invariant so that $\psi_L \to e^{i\theta}\psi_L$ and $\psi_R \to e^{i\theta}\psi_R$. The current associated with this symmetry is $J^\mu = \overline{\psi}\gamma^\mu\psi$.

For $m = 0$, one has chiral symmetry where $\psi \to e^{i\phi\gamma^5}\psi$. The axial current $J^{5\mu} = \overline{\psi}\gamma^\mu\gamma^5\psi$ is conserved, but now $\psi_L \to e^{-i\phi}\psi_L$ and $\psi_R \to e^{i\phi}\psi_R$. Note the difference in sign in the exponent indicating that the phase direction of rotation in the complex plane is opposite, and that the difference in direction corresponds to complex conjugation. What this is saying is that when one rotates a fermion, its wave function is shifted in phase in a direction that depends on the fermion's chirality. This can be seen in Figure (2.1).

The last term in the Lagrangian of Eq. (2.3) must describe how the leptons interact with the scalar Higgs field so as to make them massive. There is no theory governing this process so the form of the interaction is put in by hand. It comes from introducing a Yukawa like coupling of the scalars to leptons and further assumes that the Higgs field is a weak isotopic-spin doublet. The resulting interaction Lagrangian for the electron is

$$\mathcal{L}_{int} = G_e(\overline{\psi}_L \phi \psi_R + \overline{\psi}_R \phi^\dagger \psi_L),$$

(2.4)

where $G_e$ is an unknown coupling constant and $\phi$ is a scalar Higgs field. The form of this interaction Lagrangian was chosen to guarantee that it would be symmetric under $SU(2)_L \otimes U(1)$ and transform appropriately under Lorentz transformations. This form of interaction Hamiltonian will also maintain the zero mass of the photon and neutrino.

It is worth expanding on how mass is gained via the Higgs field, and an introduction to spontaneous electroweak symmetry breaking is given in Appendix 3.

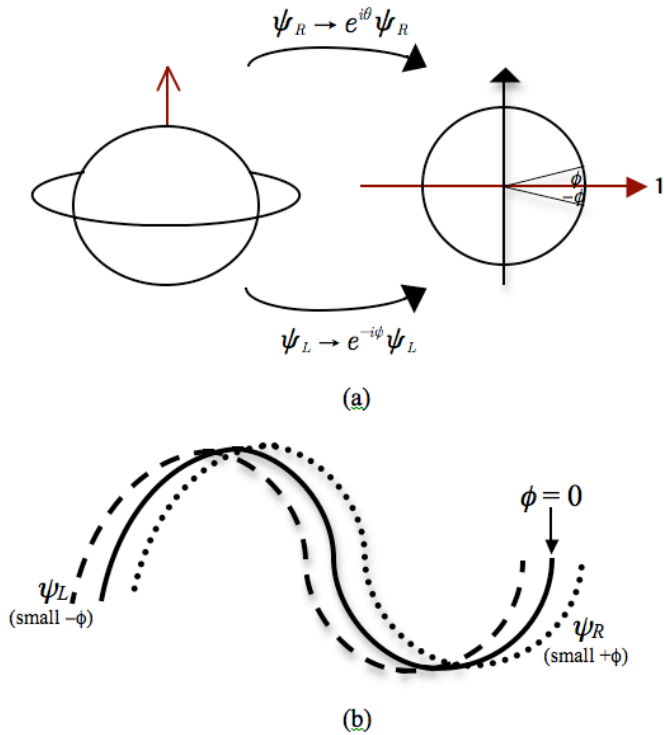

Figure (2.1). (a) When one rotates a chiral fermion about its direction of motion, both the left and right chiral fermion change be a factor of –1, but along opposite paths in the complex plane; (b) the phase shift of the particle's wave function depends on it chirality.

### *Chirality and Chiral Oscillation*

For convenience, the terms "particle" and "antiparticle"—which usually apply to a point particle—will be freely used in this section since no acceptable alternatives are readily available. These "particles" should nevertheless not be thought of as point particles. An alternative interpretation of what they might be will be given in Section 3. The electron will be used as an example and $\phi_L$ and $\phi_R$ will be designated as $e_L$ and $e_R$.

When $m = 0$, the Dirac spinors for the electron and the positron become linearly dependent so that they form a 2-dimensional vector space rather than a 4-dimensional space. The meaning of $e_L$ and $e_R$ corresponding respectively to the massive electron and positron, changes in a rather complicated way:[12] $e_L$ describes both a left-handed particle and a right handed-antiparticle and $e_R$ describes both a right-handed particle and a left-handed antiparticle. The massive particle that propagates through space is a quantum mechanical mixture of these particles and antiparticles and the mixture gains mass via interaction with the Higgs boson. This is shown in graphic form in Fig. (2.2). In the usual scenario of a cooling early universe, these particles remain massless until electroweak symmetry breaking and interaction with the Higgs boson. The Higgs field then takes a constant value everywhere.

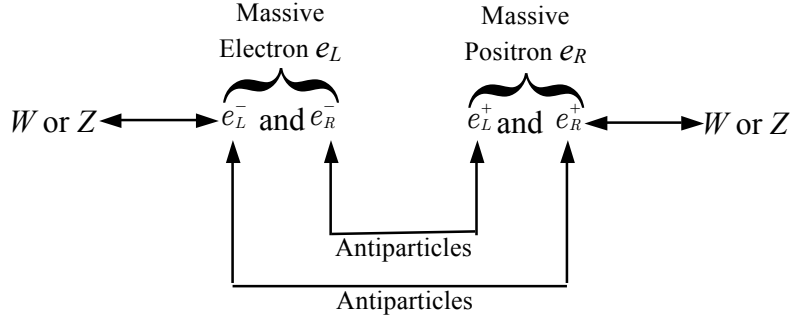

Figure (2.2). The *W* or *Z* bosons that mediate the weak force only interact with left-chiral electrons, $e_L^-$, and the right-chiral positrons, $e_R^+$. They, like the other particles shown in the figure, except for $e_L$ and $e_R$ are massless. The massive particles $e_L$ and $e_R$ are a quantum mechanical mixture of the pairs shown and appear when the mixture interacts with the Higgs non-zero vacuum expectation value. The Higgs induced mass term connects the massless left-chiral electron $e_L^-$ to the right-chiral electron $e_R^-$ and the left-chiral positron to the right-chiral positron. The right-chiral electron is designated by $e_R^-$, and the left-chiral positron by $e_L^+$. The *W* and *Z* are also massless before electroweak symmetry breaking.

It will be remembered that the eigenvalues of the velocity operator ***α*** in the Dirac equation are $\pm c$. For $m = 0$, the massless particles corresponding to $e_L$ and $e_R$ propagate at the velocity of light, but after electroweak symmetry breaking the physical electron or positron gains mass via the Higgs mechanism described in the Appendix and travels at a velocity that is always less than that of light. The combinations of these massless particles that make up the massive particles after interaction with the non-zero vacuum expectation value of the Higgs are also shown in Fig. (2.2). The *W* or *Z* bosons that mediate the weak force will only interact with left-chiral electrons and right-chiral positrons and do not interact with right-chiral electrons or left-chiral positrons. This is a consequence of the non-invariance of the weak interactions under mirror reflection symmetry induced by the parity or *P* operator.

Ordinarily, the weak interactions are invariant under the combination of charge conjugation and parity (*CP*), one exception being the decay of the $K^0$ meson.[13] In Fig. (2.2), the combination of *CP* would convert $e_L^-$ to $e_R^+$.

The left-chiral electrons, $e_L^-$, and the right-chiral positrons, $e_R^+$, could not mix to form the physical electron because they have different charges. On the other hand, the left-chiral electrons, $e_L^-$, carry a weak charge, but the right-chiral electrons, $e_R^-$, do not. One might think the interaction between them would be forbidden by gauge invariance. This is not the case because of the non-zero vacuum expectation value of the Higgs boson (which itself carries a weak charge). It is this non-zero expectation value that breaks the conservation of weak charge and allows the mixing [see Eq. (A3.7) in Appendix 3].

*Chiral Oscillations and Frequency*

In this section, the conventions and notation of De Leo and Rotelli[14] will generally be used. The chiral operator $\gamma^5$ does not commute with Dirac free-particle Hamiltonian $H = -i\vec{\alpha}\cdot\vec{\partial} + m\beta$ since $\gamma^0 = \beta$ and $[\gamma^0, \gamma^5] = 2\gamma^0\gamma^5$. Using the average value of the time derivative of an operator in the Heisenberg representation, the time derivative of the chiral operator is $\partial_t\langle\gamma^5\rangle = i\langle[H,\gamma^5]\rangle = 2mi\langle\gamma^0\gamma^5\rangle$. Note that $\partial_t\langle\gamma^5\rangle = 0$ if $m = 0$. Now,

$$\langle\gamma^0\gamma^5\rangle = \int d^3x\,\bar{\psi}(x)\gamma^5\psi(x),$$

(2.5)

where $\bar{\psi} = \psi^\dagger\gamma^0$ is the Dirac adjoint so that $\bar{\psi}\psi = \psi^\dagger\gamma^0\psi = |\psi_1|^2 + |\psi_2|^2 - |\psi_3|^2 - |\psi_4|^2$ is an invariant quantity. If $E = k^0 = (\vec{k}^2 + m^2)^{1/2}$ and the Dirac spinors are normalized as

$$v_\alpha^\dagger(k)v_\beta(k) = u_\alpha^\dagger(k)u_\beta(k) = \frac{E}{m}\delta_{\alpha\beta}$$
$$\bar{u}_\alpha(k)u_\beta(k) = -\bar{v}_\alpha(k)v_\beta(k) = \delta_{\alpha\beta}$$
$$\int d^3x\,\psi^\dagger(x)\psi(x) = 1,$$

(2.6)

$\Psi(x)$ can be expanded in terms of plane waves as

$$\psi(x) = \int\frac{d^3k}{(2\pi)^3}\frac{m}{E}\sum_{\alpha=1}^{2}[a_\alpha(k)u_\alpha(k)e^{-ikx} + b_\alpha^*(k)v_\alpha(k)e^{ikx}].$$

(2.7)

Note that the time dependence is contained in $ikx$ since $k$ and $x$ are 4-vectors. The coefficients $a_\alpha(k)$ and $b_\alpha(k)$ must satisfy

$$\int\frac{d^3k}{(2\pi)^3}\frac{m}{E}\sum_{\alpha=1}^{2}[|a_\alpha(k)|^2 + |b_\alpha(k)|^2] = 1.$$

(2.8)

The evaluation of Eq. (2.5) is tedious, and results in

$$\langle\gamma^0\gamma^5\rangle = \int\frac{d^3k}{(2\pi)^3}\frac{m^2}{E^2}\sum_{\alpha,\beta}[b_\alpha(\tilde{k})a_\beta(k)\bar{v}_\alpha(\tilde{k})\gamma^5 u_\beta(k)e^{-2iEt} - h.c.],$$

(2.9)

where $\tilde{k} = (E, -\vec{k})$ and $h.c.$ stands for "hermitian conjugate". All that is really needed to illustrate the chiral oscillation are the cross terms given by $\bar{v}_\alpha(\tilde{k})\gamma^5 u_\beta(k)$. It can be shown that this term results in the non-zero quantity

$$\bar{v}_\alpha(\tilde{k})\gamma^5 u_\beta(k) = -\frac{E}{m}\bar{v}_\alpha(\tilde{k})\gamma^0\gamma^5 u_\beta(k).$$

(2.10)

This non-zero term shows that the chiral oscillation does not vanish and can be seen from Eq. (2.9) to have the frequency $2mc^2/\hbar$, which is identical to the zitterbewegung frequency. As is the case for the zitterbewegung, for $\psi(x)$ composed of only positive or negative frequencies, $\langle \gamma^0 \gamma^5 \rangle = 0$.

The chiral representation has been used above. Switching now to the standard representation for the Dirac matrices (following De Leo and Rotelli), one can show that $-\gamma^0 \gamma^5 u_\beta(k) = v_\beta(\tilde{k})$ so that Eq. (2.10) [using the normalization given in Eqs. (2.6)] becomes

$$\bar{v}_\alpha(\tilde{k}) \gamma^5 u_\beta(k) = -\frac{E}{m} \bar{v}_\alpha(\tilde{k}) \gamma^0 \gamma^5 u_\beta(k) = \frac{E}{m} \bar{v}_\alpha(\tilde{k}) v_\beta(\tilde{k}) = -\frac{E}{m} \delta_{\alpha\beta}. \tag{2.11}$$

Now, the time derivative of the chiral operator was shown above to be $\partial_t \langle \gamma^5 \rangle = 2mi \langle \gamma^0 \gamma^5 \rangle$ and $\langle \gamma^0 \gamma^5 \rangle$ was given in Eq. (2.9). One can then write $\partial_t \langle \gamma^5 \rangle$ as

$$\partial_t \langle \gamma^5 \rangle = \int \frac{d^3k}{(2\pi)^3} \frac{m^2}{E^2} \sum_\alpha [-2iE\, b_\alpha(k)\, a_\alpha(\tilde{k})\, e^{-2iEt} + h.c.]. \tag{2.12}$$

If this is now integrated between 0 and $t$, the result is

$$\langle \gamma^5 \rangle(t) = \langle \gamma^5 \rangle(0) + \int \frac{d^3k}{(2\pi)^3} \frac{m^2}{E^2} \sum_\alpha [b_\alpha(k)\, a_\alpha(\tilde{k})(e^{-2iEt} - 1) + h.c.]. \tag{2.13}$$

Let us return now to Penrose's zig-zag interpretation of the fundamental particles of matter (the leptons and quarks), fermions having a spin of ½. While very attractive, the zig-zag model still has at its heart the concept of the point particle. An alternative will be given in the section below titled *A Topological Alternative for Charge*.

As observed by Halzen, et al.[15], ". . . the same Higgs doublet that generates the $W^\pm$ and $Z$ masses is also sufficient to give masses to the leptons and quarks". When the Higgs field has a non-zero expectation value it breaks the conservation of weak charge and allows $e_L^-$ and $e_R^-$ of Fig. 2.2 to mix so as to form the physical electron. Because the Higgs field can interact with both the $e_L^-$ and $e_R^-$ it forces them to change back and forth into each other in the chiral oscillation discussed earlier. During this oscillation, $e_L^-$ and $e_R^-$ lose their independent identity and are transformed into a single massive elementary particle. The same is true for the massless $e_L^+$ and $e_R^+$ that trough interaction with the Higgs becomes the massive positron.

In the section on photon zitterbewegung in Appendix 1, it is shown that even a non-localizable "particle" like the photon can exhibit this behavior, which is usually only associated with what historically has been called a point particle. The photon can, however, be localized for a brief moment by an interaction to about its "Compton radius", which corresponds to its classical wavelength; i.e., by assuming $E = h\nu$ and $m = E/c^2$, one has $\lambda_{c\text{-}ph} = h/mc = hc/E = c/\nu$.

The frequency of the zitterbewegung associated with the position operator for the electron is identical with that of the chiral oscillation related to $\gamma^5$. Both oscillations vanish if only positive or negative frequency solutions to the Dirac equation are used. This strongly suggests that zitterbewegung and chiral oscillations are intimately related despite their very different interpretations.

In the zig-zag concept of the electron, $e_L^-$ and $e_R^-$ continually convert themselves into the other due to their interaction with the Higgs at the chiral oscillation frequency $\nu_{ch}$, which is the same as the zitterbewegung frequency. This means the mass $m$ in Eq. (2.1), which serves as a coupling constant between these two equations, is being interpreted as a field, to quote Penrose, "essentially the Higgs field"; an illustration of this idea is shown in Fig. 2.3.

Being massless particles, the wavefunction for $e_L^-$ and $e_R^-$ after a localizing interaction expands at the velocity of light until the next localizing interaction with the Higgs. This corresponds to a distance $c/\nu_{ch} = \sim 1.3 \times 10^{-12}$ m, the Compton wavelength of the electron. Another way of looking at this is to realize that the exact time of the interaction with the Higgs is associated with a time uncertainty of $\Delta t$ and since mass before the interaction is zero and after is $m_0 c^2$, the uncertainty relation tells us that $\Delta x = c \Delta t = c\hbar/m_0 c^2 = \hbar/m_0 c = \lambda_c$, again the Compton wavelength. This is also shown in Fig. 2.3.

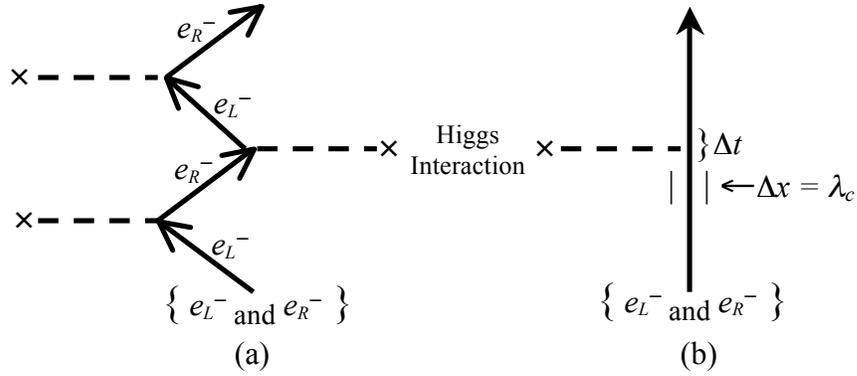

Figure 2.3. (a) $e_L^-$ and $e_R^-$ on their own are massless, but the quantum mechanical mixture acquires mass through interaction with the non-zero vacuum expectation value of the Higgs. (b) at each interaction with the Higgs field, there is an uncertainty $\Delta t$ in the time of interaction leading to a spatial uncertainty of $\Delta x = \lambda_c$.

In the Penrose model, the particles $e_L$ and $e_R$ obey the Dirac equation for a particle of mass $m$, written in the chiral form [Eqs. (2.1)] as,

$$(p_0 + \boldsymbol{\sigma} \cdot \boldsymbol{p}) e_L(\boldsymbol{p}) = m\, e_R(\boldsymbol{p}),$$
$$(p_0 - \boldsymbol{\sigma} \cdot \boldsymbol{p}) e_R(\boldsymbol{p}) = m\, e_L(\boldsymbol{p}).$$

Between Higgs interactions, the massless particles $e_L^-$ and $e_R^-$ obey the same equations with *m* equal to zero. In Fig. 2.3(a), starting with $e_L^-$ at the bottom, the first Higgs interaction destroys the left-handed massless particle $e_L^-$ and creates the right-handed antiparticle $e_R^-$ ($e_R^-$ is the right handed antiparticle of $e_L^+$). The second interaction with the Higgs field destroys the right-handed massless antiparticle $e_R^-$ and creates the left-handed massless particle $e_L^-$. It is this oscillation that is the source of the mass through the relation $E = h\nu$; i.e., $m = (1/2c^2)\, h\nu_{ch}$.

## 3. Beyond Democritus

Interpreting the absolute value of the wave function in quantum mechanics as the probability of interacting with a fundamental "particle" such as an electron at a given location does not mandate that the electron is a point particle. As put by Max Born who gave us this interpretation, ". . . we are not justified in concluding that the 'thing' under examination can actually be described as a particle in the usual sense of the term. . . . The ultimate origin of the difficulty lies in the fact (or philosophical principle) that we are compelled to use the words of common language when we wish to describe a phenomenon, not by logical or mathematical analysis, but by a picture appealing to the imagination."[16]

The Dirac or Schrödinger equations do not require that their wave functions describe the motion of a point particle. What the mathematics of quantum mechanics tells us is that an "elementary particle" is not a "particle" in the sense of classical physics. The advent of quantum mechanics mandated that the classical notion of a particle be given up. But rather than accept this, there were many attempts in the 20[th] century to retain the idea of a classical particle by a mix of classical and quantum mechanical concepts. Perhaps the best was David Bohm's 1952 theory that introduced the idea of a "quantum potential", which was later show to be equivalent to the usual quantum mechanics. None of these were really successful. In the end, we must live with the fact that an elementary particle is some form of space-time excitation that can be localized through interactions and even when not localized obeys all the relevant conservation rules and retains the "particle" properties like mass, spin, and charge.

This should not be terribly surprising since Newton-Wigner[17] and Pryce[18] showed (see discussion in Appendix 2) that a particle with spin cannot be localized to better than its Compton radius. Yet, quantum electrodynamics assumes that the electron is a point particle and electron-positron colliding beam experiments show this holds down to distances less than $10^{-18}$ *m*. Even though this is the case, it is straightforward to show that the electric charge of an electron cannot be a classical charge distribution of this or similar size that interacts with itself. If it were, its classical self-energy would exceed the rest mass energy of the electron. Another example of how classical concepts should not be carried over into quantum mechanics. Quantum electrodynamics also tells us that the effective charge of a point particle is spread out over a distance on the order of the Compton radius. This is apparent when the phenomenon of vacuum polarization is taken into account during charge renormalization.

Modern physics is telling us is that space-time can support a variety of excitations that make up the various "particles" of matter whether short lived or stable. The basic building blocks are the leptons and quarks, all of which are fermions obeying the Dirac equation for spin ½. All are associated with chiral oscillations and become massive by interaction with the spatially isotropic non-vanishing Higgs vacuum expectation value. In the Penrose model, these oscillations are continually localized by interactions with the Higgs field, and none of them need be or should be thought of as point particles. For the electron, the minimum localization is the Compton wavelength, but determining the minimum localization for quarks is more complicated than simply using the formula for the Compton wavelength because of the quark's color interactions. Estimates for the radius of the "dressed quark" divided by the radius of the proton are in the range of 0.2 to 0.5.

Perhaps the most counter-intuitive part of replacing the classical idea of a particle is accepting that a space-time excitation corresponding to a stable particle can itself carry charge, spin, angular momentum, and mass. But that is what the mathematics and physics is telling us. Such an excitation, say for an electron, is continually localized to its Compton wavelength as in Fig. 2.3, which gives it its particle aspects. It can carry spin angular momentum without being a classical particle, as does the photon. But it is perhaps charge that is the most difficult to understand.

*A Topological Alternative for Charge*

Over sixty years ago Wheeler formulated a classical explanation for charge based on electric flux threading a general relativistic "wormhole", which would doubly connect spacetime—what he called "charge without charge".[19] Einstein's field equations allow such wormholes. The concept is illustrated in Fig. 3.1.

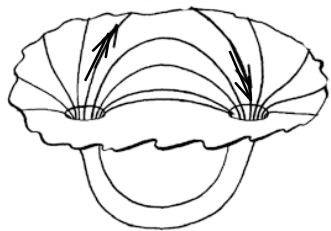

Figure 3.1. Wheeler's "charge without charge" concept where 3-dimensional space is represented as a 2-dimensional hypersurface with a wormhole. The dimension off the surface has no physical meaning. The lines threading the wormhole are the electric field lines. [Adapted from J.A. Wheeler, *Phys. Rev.* **97**, 511 (1955)]

A modern version of this idea comes from string theory[20] where a string, whose length is generally thought to be comparable to the Planck length $\sim 10^{-35}$m, terminates on a D-brane (defined below). The lowest vibrational modes of such strings are used to represent the fundamental particles of the standard model. In fundamental string theory, the strings are generally taken to be infinitely thin, although this need not be the case. The string's charge density, because of its vectorial nature, is often interpreted as a current flowing in the string, which on the brane itself is carried by the electric field. This possibility is very attractive and a basic introduction to string theory with a possible extension will be given here.

The general term "D-brane" refers to an "object" upon which, for our purposes, string endpoints lie. The letter D stands for the Dirichlet boundry conditions the endpoint must satisfy on the brane. A D$p$-brane is an object with $p$ spatial dimensions. The general spacetime dimension is $p + 1$. So 4-dimensional spacetime is considered to be a D3-brane. An example of a D2-brane with a string having endpoints on the brane is shown in Fig. 3.2. The string is drawn so as to be orthogonal to the brane. Branes with D spatial dimensions are also called D-branes. D-branes are not necessarily hypersurfaces or of infinite extent, they can also be finite, closed surfaces. The additional spatial dimensions beyond the dimension of the brane are know as comprising the "bulk".

It should be mentioned that for representing particles that are fermions, one must introduce "superstrings". The "world-sheet" of an open string is defined as the trajectory of the string in space-time with space-like coordinates $X^\mu$. On this world sheet there are two linearly independent tangent vectors given by $\partial_\tau X^\mu$ and $\partial_\sigma X^\mu$, where $\tau$ parameterizes time and $\sigma$ parameterizes the distance along the string. For bosonic strings, one uses the classical variable $X^\mu(\tau,\sigma)$ to describe the position of the string. For superstrings, the classical anti-commuting variables $\psi_\alpha^\mu(\tau,\sigma)$, ($\alpha = 1,2$), are used. Their quantization results in particle states that represent spacetime fermions. This is one of the reasons that supersymmetry is so attractive; the principal other one being that its confirmation would allow the strengths of the strong, weak, and electromagnetic forces to merge at ~$10^{15}$ GeV.

To see how string theory might offer a modern incarnation of Wheeler's "charge without charge" we need to introduce the Kalb-Ramond massless antisymmetric gauge field $B_{\mu\nu}$ = $-B_{\nu\mu}$, which is the analog of the Maxwell gauge field $A_\mu$ of electromagnetics. In the case of electromagnetics, the field strength is given by $F_{\mu\nu} = \partial_\mu A_\nu - \partial_\nu A_\mu$. For $B_{\mu\nu}$ the field strength, $H_{\mu\nu\rho}$, is defined as $H_{\mu\nu\rho} = \partial_\mu B_{\nu\rho} + \partial_\nu B_{\rho\mu} + \partial_\rho B_{\mu\nu}$. It is generally thought that the endpoints of the string shown in Fig. 3.2 have a Maxwell electric charge.

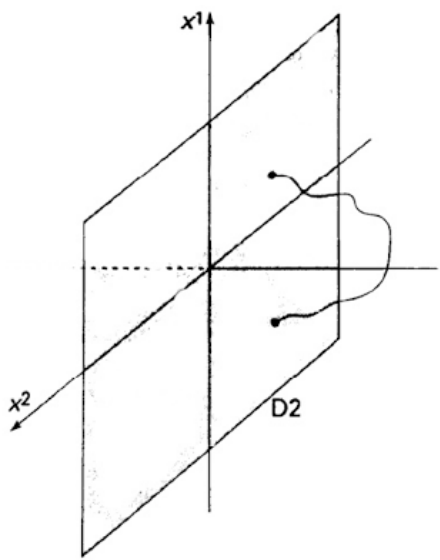

Figure 3.2. The D2-brane extends over the $(x_1, x_2)$-plane. The endpoints of the string are free to move over the plane. The Dirichlet boundary condition for the string is that the endpoint of the string cannot move out of the plane. The dimension on the unmarked axis outside the plane should *not* be thought of as a third spatial dimension belonging to the D2-brane.

The string, on the other hand, carries electric Kalb-Ramond charge. This charge can be viewed as a "current" flowing along the string; the string charge density *vector* is tangent to the string. The action for the brane and the string will have a $F^{0k}B_{0k}$ term. Since $F^{0k}$ couples to $B_{0k}$ it must carry a string charge, but $F^{0k} = E_k$, so that the Maxwell electric field on the brane carries string charge.

As mentioned before, it has been assumed here that the string is orthogonal to the brane. This is because for a string ending on a D$p$-brane, where $p \geq 2$, the velocity can be freely chosen if the string is orthogonal to the brane, whereas if it is not it must move at the velocity of light transverse to the string. A good discussion of boundary conditions and the history of the discovery of branes has been given by Tong.[21]

To summarize the overall picture given thus far, the ends of the string in Fig. 3.2 or Fig. 3.3 behave as point charges in Maxwell electromagnetics; the electric field in, for example, a D3-brane also carry string charge. The string charge on the string is a vector quantity and is analogous to a Maxwell current along the string. This is shown in Fig. 3.3 with one spatial dimension suppressed. That the endpoints of the string correspond to point particles is consistent with the experimental observation, using electron-positron colliding beams, that the electron appears as a point particle down to distances less than $10^{-18}\ m$.

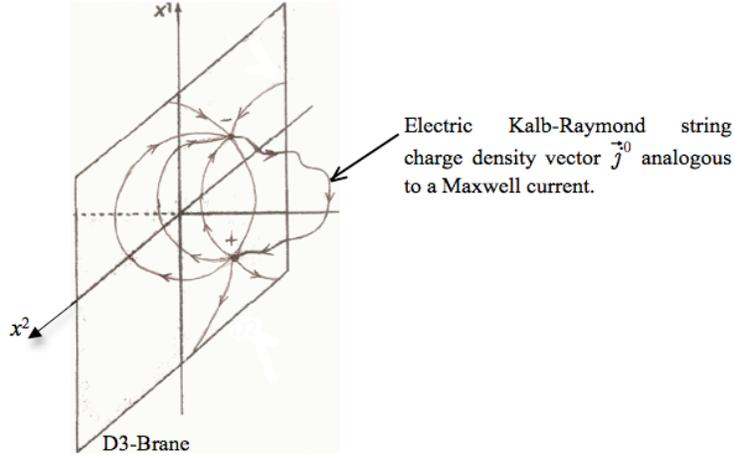

Figure 3.3. A string beginning and ending on a D3-Brane. The ends of the string behave as Maxwell point charges; the electric field lines in the D3-brane also carry string charge. The Kalb-Raymond string charge density is analogous to a Maxwell current along the string.

The following discussion explains how a Maxwell charge arises when strings terminate on branes. It comes from the fact that the conservation of string charge fails at the string endpoints. The way string theory solves this issue is to add to the string action two terms that couple the Maxwell gauge field $A_\mu$ to the Kalb-Ramond gauge field $B_{\mu\nu}$. This is done so as to preserve gauge invariance.

The field strength, $H_{\mu\nu\rho}$, defined above, is totally antisymmetric and invariant under the gauge transformations

$$\delta B_{\mu\nu} = \partial_\mu \Lambda_\nu - \partial_\nu \Lambda_\mu.  \tag{3.1}$$

Here the arguments of $B_{\mu\nu}$ are the string coordinates $X(\tau, \sigma)$.

The part of the action that couples the string to the $B_{\mu\nu}$ field is given by

$$S_B = -\frac{1}{2} \int d\tau d\sigma \, \varepsilon^{\alpha\beta} \partial_\alpha X^\mu \partial_\beta X^\nu B_{\mu\nu}.  \tag{3.2}$$

$\varepsilon^{\alpha\beta}$ is totally antisymmetric so that when this action is varied using Eq. (3.1) the result is

$$\delta S_B = -\int d\tau d\sigma \left(\partial_\tau \Lambda_\nu \, \partial_\sigma X^\nu - \partial_\sigma \Lambda_\nu \, \partial_\tau X^\nu \right) = -\int d\tau d\sigma \left(\partial_\tau (\Lambda_\nu \partial_\sigma X^\nu) - \partial_\sigma (\Lambda_\nu \partial_\tau X^\nu)\right).  \tag{3.3}$$

The second equality is a result of $\partial_\sigma X^\nu$ not being a function of $\tau$ and $\partial_\tau X^\nu$ not being a function of $\sigma$.

Now if $\Lambda$ is set equal to zero at $\pm\infty$, the $\partial_\tau$ term vanishes. Since the string terminates on a D-brane, the coordinates $X^\mu$ may be divided into those on the brane labeled $X^m$ and those perpendicular to the brane labeled $X^a$. Then integrating Eq. (3.3) on the brane with respect to $\sigma$ gives

$$\delta S_B = \int d\tau \, (\Lambda_m \partial_\tau X^m + \Lambda_a \partial_\tau X^a) \Big|_{\sigma=0}^{\sigma=\pi} .$$
(3.4)

Because Dirichlet boundary conditions apply at both end points of the string, the term $\Lambda_a \partial_\tau X^a$ vanishes when evaluated at these points.

For $S_B$ to be gauge invariant $\delta S_B$ must vanish. To make this happen one adds a term to the action coupling the ends of the string to the Maxwell fields on the brane. That is,

$$S = S_B + \int d\tau \, A_m(X) \partial_\tau X^m \Big|_{\sigma=0}^{\sigma=\pi} .$$
(3.5)

For this to work, one must impose the condition $\delta A_m = -\Lambda_m$. Doing so immediately results in $\delta S = 0$ so that gauge invariance is restored. Now, however, since $\delta F_{mn} = -\delta B_{mn}$ neither field is independently gauge invariant. This means that the physical field strength must be redefined as $\mathcal{F}_{mn} = F_{mn} + B_{mn}$. Then on the brane the gauge invariant generalization of the Maxwell Lagrangian density is $-\frac{1}{4}\mathcal{F}^{mn}\mathcal{F}_{mn}$. Expanding this gives

$$-\frac{1}{4}\mathcal{F}^{mn}\mathcal{F}_{mn} = -\frac{1}{4}B^{mn}B_{mn} - \frac{1}{4}F^{mn}F_{mn} - \frac{1}{2}F^{mn}B_{mn} .$$
(3.6)

The last term can be written as

$$-\frac{1}{2}F^{mn}B_{mn} = -F^{0k}B_{0k} + \ldots$$
(3.7)

Since $F^{0k}$ couples to $B_{0k}$ it must carry a string charge, but $F^{0k} = E_k$ so that Maxwell electric field on the brane carries string charge.

The real question is how to interpret the second term of Eq. (3.5),

$$\int d\tau \, A_m(X) \partial_\tau X^m \Big|_{\sigma=\pi} - \int d\tau \, A_m(X) \partial_\tau X^m \Big|_{\sigma=0} .$$
(3.8)

It is generally maintained that these terms add a plus and minus Maxwell charge to the ends of the string. But the first term on the right hand side of Eq. (3.7) can be interpreted as saying that not only does the Maxwell electric field on the brane carry string charge, but the string in the "bulk" carries the electric field as well. This is what the $F^{0k}B_{0k}$ term in Eq. (3.7) means—the two fields are coupled. If this is the case, there need be no charge at the terminations of the string on the brane but just the emergence of the field lines, which would look like charges within the brane; essentially as shown in Fig. 3.3 with the + and – symbols corresponding to the entering and leaving of the field lines in the brane rather than charges. Gauge invariance is conserved since there is no longer a boundary for $\sigma$ and the $\partial_\sigma$ term in Eq. (3.3) vanishes.

Using two parallel branes one can give a modern version of Wheeler's "charge without charge". The use of two is important for if only one were used (as in Fig. 3.3) with a string having both ends attached to it to represent particles with opposite charge, the motion of the charges could possibly affect string tension and hence mass. With two branes having constant separation the motion of the particles need not affect the string tension provided the motions of the string ends on each brane mirror each other. The configuration is shown in Fig. 3.4.

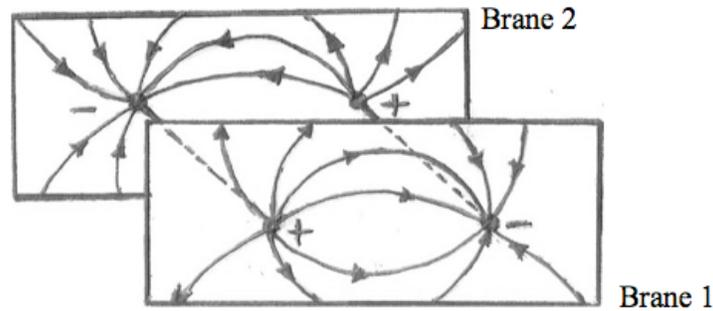

Figure 3.4. Two brane version of Wheeler's "charge without charge". The dashed lines correspond to strings that carry both the Electric Kalb-Raymond string charge density vector $\vec{j}^0$ and the Maxwell electric field. The same is true of the Maxwell field lines within the two branes as suggested by the coupling $F^{0k}B_{0k}$. The dots at the end of the strings indicate where the fields enter (+) or leave (–) the branes—not actual charges. The strings move in tandem with the oppositely "charged" string ends and the spacing between the branes does not change during the motion so as to keep the mass associated with the string ends constant. Motion of a string end in brane 1 is mirrored by motion of the oppositely charged string end in brane 2.

There is no necessity to associate charges with the string ends since neither the Kalb-Raymond string charge density vector or Maxwell electric field terminates there. But it remains to discuss the nature of the string itself.

The trajectory of a zero-dimensional point particle is a one-dimensional curve. In string theory a string is often taken to be the same thing—a one-dimensional space. Since the lowest vibrational modes of such strings are used to represent the fundamental particles of the standard model, if the string terminates on a brane, the oscillations within the brane of the string endpoint could be identified with the zitterbewegung of the particle. So long as the string diameter is zero, the oscillations of the end point would be equivalent to the oscillations of a point particle. There is then an equivalency between the concept of a point particle and a string with zero thickness.

Tong (Ref. 21) considers the magnetic flux tubes in Type II superconductors and the chromo-electric flux tubes in QCD to be strings and notes that there are two length scales associated with the string, the tension and the width of the string, where the tension is the energy per unit length of the string.

If the string were allowed to have a finite diameter when it terminates on a brane, the endpoint of the string would not correspond to a point particle. For an electron, the diameter would be similar to the Compton wavelength, although the field could be localized to far smaller distances. Since in the model of Fig. 3.4 the string carries the Maxwell electric field as well as the Kalb-Raymond string charge density, allowing the string to have a finite diameter limits the strength or magnitude of the Maxwell electric field in the string (intuitively, the electric field strength is proportional to the number of "field lines" divided by the area perpendicular to the field lines). From the point of view of an observer in either of the branes, the situation looks like that shown in Fig. 3.1.

An ordinary string has transverse oscillations that propagate at a velocity of $v = \sqrt{T/\mu}$, where $T$ is the tension and $\mu$ is the linear mass density. Consider again the electron. If the Compton wavelength sets the scale, because the tension is given by the energy per unit length one has $T = m_e c^2 / (h/m_e c)$. Substituting this into the expression for the velocity yields $v = c$. This is consistent with the string oscillations being identified with the zitterbewegung and with the eigenvalues of the Dirac velocity operator $\vec{\alpha}$ being $\pm c$ (see Appendix 2 on zitterbewegung). In this case, the distance between the branes in Fig. 3.4 would be the Compton wavelength $h/m_e c$.

The string theory representation of the particles of the standard model of particle physics introduces additional complications. In order to accommodate the elementary fermionic building blocks of leptons and quarks, three parallel branes that carry color (red, blue, green) are introduced as well additional branes corresponding to isospin and chirality. The three color branes correspond to SU(3)$_C$. Quarks are then open strings with one endpoint on one of these branes. Gluons have both ends terminating on one of these branes. Antiquarks correspond to oppositely oriented strings. The other ends of such open strings terminate on branes having the appropriate isospin and chirality. The fermions in the standard model require the specification of color, isospin, and hypercharge.

If strings stretch between parallel but not coincident branes their quantum fluctuations give rise to massive particles with a mass proportional to their separation because of string tension. If they are coincident the fluctuations are massless. It should be remembered that in the standard model masses are not allowed for chiral fermions before symmetry breaking. The zero mass requirement can be accommodated by having the color branes intersect the branes corresponding to different chirality and isospin. Then the fermion fields will be represented by strings localized to the intersections. The branes that intersect the color branes are know as left and right branes, which correspond to the left and right handed particles. The various sets of branes are then color branes, left and right branes (sometimes called the weak branes), and leptonic branes.

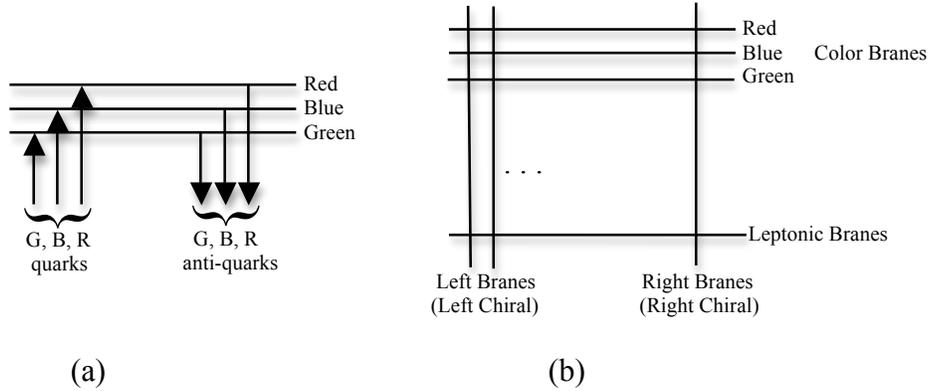

(a)                  (b)

Figure 3.5. (a) Open strings with one end on the brane. The ones on the left directed toward the brane correspond to green, blue, and red quarks; those on the right directed away from the brane are the corresponding anti-quarks. (b) some of the branes of the standard model. The various particles are appear at the crossings. For zero-mass particles the parallel branes are coincident.

All of this adds up to a graphical representation of the standard model in terms of branes. Some of the way this looks is shown in Fig. 3.5. When the branes are coincident their fluctuations are massless and mix to form a matrix valued field.

To go beyond this point, the reader is referred to the standard texts on string theory.

## *SUMMARY*

The concept of the charged point particle already had serious problems in classical physics and the earlier parts of this essay should have made it clear that the same thing is true in quantum mechanics. All indications are that the concept of a point particle loses its meaning as one approaches the Compton wavelength—notwithstanding the fact that "point particles" can be localized to less than this by high energy scattering experiments.

Wheeler's attempt to introduce the concept of "charge without charge" in the mid-1950s failed primarily because general relativity could not be quantized, which—despite enormous effort since then—remains true today. The advent of string theory may offer an alternative to the point particle despite the fact that there is no experimental evidence that string theory is more than a fascinating and beautiful mathematical exercise.

The conception of introducing additional space-like dimensions to four-dimensional spacetime coupled with the concept of strings that could exist in this enlarged space introduces the possibility of a new formulation of Wheeler's "charge without charge". To do so one must allow strings to have a width comparable to the Compton wavelength and allow them to carry an electric field just as an electric field can carry string charge. This is strongly implied by the coupling term $-F^{0k}B_{0k}$ in Eq. (3.7).

## APPENDIX I: Electron Channeling Resonance in Crystals

Consider a line of atoms, corresponding to those of a silicon crystal along the <110> direction, having a spacing $d$. Electrons traveling parallel to this line of atoms are assumed to carrying an internal clock having a frequency corresponding to the de Broglie frequency. Call the time measured by this clock $T$, while that in the laboratory is designated by $t$. The de Broglie frequency is $\omega = m_0 c^2 / \hbar$ and $\omega$ changes by $2\pi$ radians in one cycle. As the electron moves along its path, the phase velocity represents the change in $\vec{r}$ with respect to $T$; that is,

$$\frac{d\vec{r}(t)}{dT(t)} = v_{ph}.$$

(A1.1)

Then, $d / v_{ph} = \Delta T$. The phase angle change as the electron moves from one atom to the next is

$$\Delta\varphi = \Delta T\omega = \left(\frac{d}{v_{ph}}\right)\left(\frac{m_0 c^2}{\hbar}\right).$$

(A1.2)

Solving for $\Delta T$,

$$T = \frac{\Delta\varphi}{\omega} = \Delta\varphi\left(\frac{\hbar}{m_0 c^2}\right).$$

(A1.3)

Now set $\Delta\varphi = n\pi$ (note that odd $n$ only changes the sign),

$$(\Delta T)_n = n\pi\left(\frac{\hbar}{m_0 c^2}\right) = \frac{n}{2}\left(\frac{h}{m_0 c^2}\right).$$

(A1.4)

This is $n/2$ times the de Broglie period. From $\Delta T = d / v_{ph}$,

$$(v_{ph})_n = d\left(\frac{n}{2}\right)\left(\frac{m_0 c^2}{h}\right).$$

(A1.5)

Now $v_{ph} v_g = c^2$ so that $v_{ph} = \dfrac{c^2}{v_g} = \dfrac{m_0 c^2}{m_0 v_g} = \dfrac{E}{p}$ and

$$\frac{E_n}{c} = p\left(\frac{v_{ph}}{c}\right).$$
(A1.6)

As will be shown shortly, $v_g$ = (1–1/158) so that there is little error in replacing $p = m_0 v_g$ with $m_0 c$. Also, $\gamma \sim 1$ so that there is no need to make a relativistic correction. Thus,

$$\frac{E_n}{c} = m_0 c^2 \left(\frac{v_{ph}}{c}\right)\frac{1}{c}.$$
(A1.7)

As mentioned earlier, for silicon along the <110> direction, $d = 3.84 \times 10^{-10} m$. Substituting this and the numerical values for the other symbols in Eq. (A1.5), and using the result in Eq. (A1.7), gives

$$\frac{E_n}{c} = \left(\frac{2}{n}\right)m_0 c^2 (158)\frac{1}{c}.$$
(A1.8)

If the units of energy are now chosen to be MeV, the final result for the momentum is

$$\frac{E_n}{c} = \left(\frac{2}{n}\right)0.51(158)\frac{1}{c} = \left(\frac{2}{n}\right)80.58 Mev/c.$$
(A1.9)

The data shown in Fig. 1.1 imply that $n = 2$. One can obtain the zitterbewegung frequency rather than the de Broglie frequency by using $d/2$ rather than $d$ in the expressions above or by assuming $n = 1$, but at this point there is no strong experimental reason for doing either.

So regardless of whatever discrepancies may exist between modeling results and the experimental data, there would appear to be good reason to believe de Broglie's supposition that each particle carries with it a kind of "internal clock", which had previously been verified only indirectly by the many successful tested implications of the de Broglie relation.

## APPENDIX 2: Zitterbewegung

The first part of this section discusses the zitterbewegung of particles obeying the Dirac equation, and the second part shows that the photon also exhibits zitterbewegung. The purpose of showing that this is the case is to disassociate zitterbewegung from the conception of a point particle and show that it is even exhibited by the familiar electromagnetic wave.

With $c = 1$, the Dirac Hamiltonian is $H = \vec{\alpha} \cdot \vec{p} + \beta m$, where $\vec{\alpha}$ is the velocity $\vec{\alpha} = d\vec{r}/dt$ and $\beta$ is four by four matrix $\beta = \begin{pmatrix} 1 & 0 \\ 0 & -1 \end{pmatrix}$.

In the Heisenberg representation, the time derivative of an operator is given by $\frac{dO}{dt} = i[H, O] + \frac{\partial O}{dt}$, so that for the position operator $\vec{r}$,

$$\frac{d\vec{r}}{dt} = i[H, \vec{r}] = i\vec{\alpha}[\vec{p}, \vec{r}] = \vec{\alpha}.$$

(A2.1)

Since $\vec{\alpha}$ only acts on spin variables, this defines each component of $\dot{\vec{r}}$ as a constant matrix. The time derivative of $\vec{\alpha}$ is

$$\frac{d\vec{\alpha}}{dt} = i[H, \vec{\alpha}] = i(H\vec{\alpha} + \vec{\alpha}H) - 2i\vec{\alpha}H.$$

(A2.2)

Now if one expands the first term on the r.h.s. of the latter equation in terms of $\vec{\alpha} = (\alpha_1, \alpha_2, \alpha_3)$ and $\vec{p} = (p_1, p_2, p_3)$, the result is

$$\frac{d\vec{\alpha}}{dt} = 2i\vec{p} - 2i\vec{\alpha}H.$$

(A2.3)

One can solve this equation for $\vec{\alpha}$ by integrating from 0 to $t$ and, remembering that $\vec{\alpha} = d\vec{r}/dt$, integrate again between the same limits to obtain the solution for $\vec{r}$, yielding

$$\vec{r}(t) = \vec{r}(0) + \frac{\vec{p}}{H}t + i\left(\vec{\alpha}(0) - \frac{\vec{p}}{H}\right)\frac{e^{-2iHt}}{2H}.$$

(A2.4)

The term $e^{-2iHt}$ corresponds to a circular motion since it can be written as $(\cos 2H - i\sin 2H)$. At rest, $H = mc^2$ and writing $e^{-2iHt}$ as $e^{-i\omega t}$, and putting in the $\hbar$, which has been set equal to unity in this calculation, gives $\omega = 2mc^2/\hbar$. This is the frequency of the zitterbewegung.

The classical law for uniform rectilinear motion is given by the first two terms on the right hand side when they are not operators. It is the last term that is responsible for the zitterbewegung, and it is usually interpreted as meaning that the particle samples a region on the order of its Compton wavelength, $\delta r \approx \hbar/mc$, about the point $\vec{r}$. Zitterbewegung is generally thought to be due to interference between negative and positive frequency states, as was originally proposed by Schrödinger.

It is at this point that the interpretation of fermions, such as electrons and quarks, as point particles runs into serious problems. This comes about by calculating the velocity eigenvalues of the velocity operator $\vec{\alpha}$. Putting back the constant factors of $c$ in the Dirac equation one gets $\dot{\vec{r}}(t) = c\vec{\alpha} = c\gamma^0\vec{\gamma}$. Using the explicit matrices for the $\gamma$ matrices, one obtains for the eigenvalues $\pm c$. This is usually interpreted to mean that while the average velocity of the particle may be less than $c$, the instantaneous velocity is always $\pm c$, but the meaning of this, given that the particles are massive and charged, is far from clear.

In a paper more than fifty years ago, Huang[22] used the expectation values of $\vec{r}$ and $\vec{r} \times \dot{\vec{r}}$ in a wave packet representing the electron to show that the zitterbewegung could be interpreted as a circular motion about the direction of the electron spin. The radius of the motion was $\hbar/2m$, the Compton wavelength divided by $2\pi$. The intrinsic spin was then the "orbital angular momentum" of this motion, and the current produced gives rise to the intrinsic magnetic moment. This is derived from Dirac theory and is not a classical interpretation of the results. However, implicit in his discussion is the assumption that the electron is a point particle. This is also an assumption made by many others who have attempted to formulate classical models of fermions based on the zitterbewegung phenomenon. The implication, since the eigenvalues of the velocity operator are $\pm c$, is again that a point particle carrying both mass and charge can move in a circular orbit at the velocity of light. Clearly, there is a problem.

*Photon Zitterbewegung*

It should be remembered that a photon is not a true particle in that it is not localizable—a subject that will be discussed latter in this section—but represents the minimum amount of energy that an electromagnetic wave can carry, and a wave can only carry multiples of this minimal amount. It was Einstein that introduced the idea of a photon in an attempt to deal with the wave-particle dilemma early in the history of quantum mechanics. To quote Leon Rosenfeld, Einstein made the qualitative suggestion "that the photons, or the light quanta as they were called then, were some kind of singularity, of concentration of energy and momentum inside a radiation field. The radiation field would so to speak guide the photons in such a way as to produce also the interference a diffraction phenomena . . . ."[23] The confused interpretation of a photon as a particle continues to this day. Einstein's suggestion is found in the de Broglie-Bohm interpretation of quantum mechanics.

To discuss photon zitterbewegung, one needs a wave equation for the photon. The approach that will be used to find one is based on that of Bialynicki-Birula.[24] The result will be an Schrödinger like wave equation for the photon. This will then allow the Hamiltonian for the photon to be identified.

D'Alenbert's equation

$$\frac{1}{c^2}\partial_t^2 \Phi = \Delta \Phi$$

(A2.5)

will be used a guide for finding an analogous relation for a spin-1 massless particle like the photon. Taking the square root of this equation gives

$$\frac{1}{c}\partial_t \Phi = \sqrt{\Delta}\, \Phi = \vec{\nabla} \Phi.$$

(A2.6)

The photon having spin-1 suggests that the three Hermitian matrices representing infinitesimal rotations for spin-1, $(S_i)_{kl} = -i\varepsilon_{ikl}$ be the starting point for finding a wave equation for the photon. These matrices are

$$S_x = \begin{pmatrix} 0 & 0 & 0 \\ 0 & 0 & -i \\ 0 & i & 0 \end{pmatrix}, \quad S_y = \begin{pmatrix} 0 & 0 & i \\ 0 & 0 & 0 \\ -i & 0 & 0 \end{pmatrix}, \quad S_z = \begin{pmatrix} 0 & -i & 0 \\ i & 0 & 0 \\ 0 & 0 & 0 \end{pmatrix}.$$

(A2.7)

They obey the anti-commutation relations,

$$[(S_i)(S_j) + (S_j)(S_i)]_{ab} = 2\delta_{ij}\delta_{ab} - \delta_{ai}\delta_{bj} - \delta_{aj}\delta_{bi}.$$

(A2.8)

Multiply both sides of this relation by $\nabla_i \nabla_j$, and rearrange the terms to get

$$[(\vec{S}\cdot\vec{\nabla})(\vec{S}\cdot\vec{\nabla})]_{ab} = \nabla^2 \delta_{ab} - \nabla_a \nabla_b.$$

(A2.9)

If we let this operator relation operate on $\vec{\psi}$,

$$[(\vec{S}\cdot\vec{\nabla})(\vec{S}\cdot\vec{\nabla})]_{ab}\vec{\psi} = \nabla^2 \delta_{ab}\vec{\psi} - \nabla_a \nabla_b \vec{\psi}.$$

(A2.10)

The first term on the right hand side vanishes unless $a = b$; choose this to be the case. Then when the remaining index is summed over, the second term becomes $\vec{\nabla}(\vec{\nabla}\cdot\vec{\psi})$, which—since there are no sources—will be assumed to vanish. The resulting equation is true for all *a* so the index can be dropped. Dropping the $\vec{\psi}$ from both sides leaves only the operator relation, which can be written as

$$(\vec{S}\cdot\vec{\nabla})^2 = \vec{\nabla}^2 \quad or \quad \pm(\vec{S}\cdot\vec{\nabla}) = \vec{\nabla}.$$

(A2.11)

Now multiply both sides by $\psi$ and compare the result with Eq. (A2.6). This comparison suggests that $\psi$, the wave function for the photon, be considered to be a 3-component spinor

$$\psi = \begin{pmatrix} \psi_1 \\ \psi_2 \\ \psi_3 \end{pmatrix},$$

(A2.12)

where the $\psi_i$ are scalar functions. On a component basis then, we have

$$\pm(\vec{S}\cdot\vec{\nabla})\psi_i = \vec{\nabla}\psi_i, \quad \text{which is to be compared to} \quad \frac{1}{c}\partial_t \Phi = \vec{\nabla}\Phi.$$

(A2.13)

This leads to the final classical result to be written as

$$\frac{1}{c}\partial_t \psi = \pm(\vec{S}\cdot\vec{\nabla})\psi.$$

(A2.14)

This can be converted into a quantum mechanical expression by simply multiplying both sides by $i\hbar$

$$i\hbar\partial_t \psi = \pm ci\hbar(\vec{S}\cdot\vec{\nabla})\psi = \mp c(\vec{S}\cdot[-i\hbar\vec{\nabla}])\psi = \mp c(\vec{S}\cdot\vec{p})\psi.$$

(A2.15)

The Hamiltonian for the photon is then $\mp c(\vec{S}\cdot\vec{p})$. For positive energy choose the + sign, which corresponds to +helicity. Since the $S_i$ are pure imaginary, sign reversal corresponds to complex conjugation.

Using Eq. (A2.7), the Hamiltonian in component form can be written as

$$H_{kl} = \lambda c(\vec{S}\cdot\vec{p})_{kl} = \lambda c(S_i p_i)_{kl} = \lambda c(S_i)_{kl} p_i = -i\lambda c\varepsilon_{ikl} p_i.$$

(A2.16)

To be consistent with Kobe,[25,26] relabel the indices by $k \to i, l \to k, i \to j$ giving $H_{ik} = i\lambda c\varepsilon_{ijk} p_j$.

Hamilton's equation for the velocity is $v_j = \partial H / \partial p_j$. If $v$ and $H$ are considered to be operators, one can write

$$(v_j)_{ik} = \partial_{p_j} H_{ik} = \partial_{p_j}(i\lambda c \varepsilon_{ijk} p_j) = i\lambda c \varepsilon_{ijk}.$$

(A2.17)

Thus, $H_{ik} = i\lambda c \varepsilon_{ijk} p_j$ is the same as $H = \vec{v} \cdot \vec{p}$. This is the form of the Hamiltonian that will be used to compute the zitterbewegung.

The time derivative of an operator in the Heisenberg representation, $\frac{dO}{dt} = i[H, O] + \frac{\partial O}{\partial t}$, can be used to compute the derivative with respect to time of the velocity operator,

$$\frac{d\vec{v}}{dt} = (i\hbar)^{-1}[\vec{v}, H] = \overleftrightarrow{H} \cdot \vec{v}.$$

(A2.18)

Following Kobe, the dyadic form for $H$ has been introduced to facilitate the integration of this equation when $\vec{v}$ and $\vec{p}$ are operators. Since dyadic notation is used only sparsely in the modern physics literature, the key to understanding Eq. (A2.18) is the relation

$$(\vec{A}\vec{B}) \cdot \vec{C} = \vec{A}(\vec{B} \cdot \vec{C}).$$

(A2.19)

The first term on the left hand side within the parentheses is a dyad; note that there is no operation defined between the vectors. Thus, $\overleftrightarrow{H} \cdot \vec{v} \rightarrow (\vec{v}\vec{p}) \cdot \vec{v} = \vec{v}(\vec{p} \cdot \vec{v})$, and

$$\frac{d\vec{v}}{dt} = (i\hbar)^{-1} \vec{v}(\vec{p} \cdot \vec{v}) = (i\hbar)^{-1} \vec{v} H.$$

(A2.20)

Operating on the left with $\vec{v}^{-1}$ and integrating from 0 to $t$ gives

$$\vec{v}(t) = \vec{v}(0) e^{-\frac{i}{\hbar} H t}.$$

(A2.21)

Note that while the calculation is somewhat tedious, one can explicitly show that $[\vec{v}, H] = \vec{v}(\vec{p} \cdot \vec{v})$ using the matrix definitions for $\vec{v}$ given above in Eq. (A2.17).

Now $\vec{v}$ is a constant velocity operator, which means that $\vec{v}(t)$ depends entirely on $\vec{v}(0)$. At $t = 0$, the velocity is directed along the momentum, $\vec{v}(0) = c\hat{p}$, where $\hat{p} = \vec{p}/|\vec{p}| = \vec{p}/p$. Then $\vec{v}(t)$ can be written in terms of its components parallel and perpendicular to $\hat{p}$, as $\vec{v}(t) = \vec{v}_\parallel(0) + \vec{v}_\perp(0)$. At $t = 0$, only $\vec{v}_\parallel(0)$ is not zero but at a later time this is not the case, so that in terms of the parallel and perpendicular components one can write

$$\vec{v}(t) = \vec{v}_\parallel(0) + \vec{v}_\perp(0) e^{-\frac{i}{\hbar}Ht}.$$

(A2.22)

The only condition is that $\hat{p} \cdot \vec{v}_\perp = 0$. If Eq. (A2.22) is integrated, we get what Kobe calls a "displacement operator" (it cannot be called a "position operator" since the photon is not localizable), which leads to a short digression on position operators in quantum mechanics.

Newton and Wigner and Pryce have given thorough discussions of position operators. It is the spin that is responsible for the photon's non-localizability,[27] If the photon had spin zero, it would be localizable. Newton-Wigner derive an expression for the position coordinate for arbitrary spin, but for spin ½ it agrees with Pryce who defines the center of mass in coordinates where the coordinates taken in pairs have vanishing Poisson brackets. In such a frame, the total momentum vanishes, and the center of mass is at rest—a result that is frame dependent. Note that the center of mass of a single particle is the same as the position of the particle. Pryce concludes, "From the point of view of relativistic quantum mechanics the only 'position vector' that has much interest is the one which is relativistically covariant . . . The fact that its components do not commute leads to an uncertainty in the simultaneous measurement of order $\hbar/mc$". Or, as put by Bacry, "either it is impossible to measure any coordinate, that is there is no position operator, or the position operator has three non-commuting components". In particular, massive particles with spin can be localized to a minimal uncertainty in one frame of reference, but in another frame it will not be localized—localized states are not transformed into localized states under Lorentz transformations.

Returning to the discussion of Eq. (A2.22), remembering that $\vec{v}(t) = d\vec{x}(t)/dt$ the integral of the equation is

$$\vec{x}(t) = \int_0^t \vec{v}_\parallel(0) dt + \int_0^t \vec{v}_\perp(0) e^{-\frac{i}{\hbar}Ht} dt.$$

(A2.23)

But rather than integrating this equation in its present form, it is advantageous to first determine the form of $\vec{v}_\perp(t)$.

It was shown in Eq. (A2.15) that the Hamiltonian for for the photon wave function is $\mp c(\vec{S} \cdot \vec{p})$. The positive sign was chosen corresponding to a positive helicity. What the ± sign means is that there are two independent parts of the wave function corresponding to the positive and negative states of helicity. Bialynicki-Birula introduced a 6-dimensional wave function with a single evolution equation to deal with the two helicity states of the photon. Another way to introduce helicity, following Kobe, is to write $\vec{v}_\perp(0)$ in Eq. (A2.23) in terms of two orthogonal components such that the two circular polarization vectors $\hat{e}_\varepsilon$, where $\varepsilon = \pm 1$, are given by

$$\hat{e}_\varepsilon = \frac{1}{\sqrt{2}}(\vec{v}_\perp + i\varepsilon \vec{v}'_\perp).$$

(A2.24)

The arrangement of these vectors is shown in Fig. (A2.1).

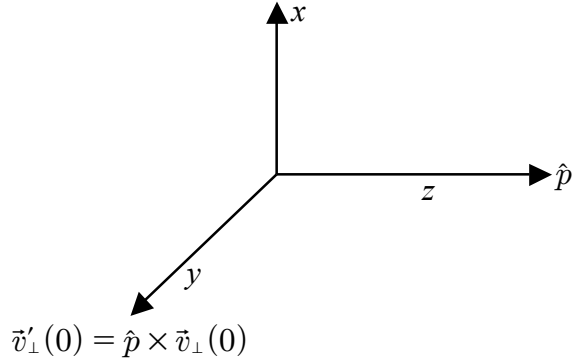

$$\vec{v}'_\perp(0) = \hat{p} \times \vec{v}_\perp(0)$$

Figure (A2.1). The arrangement of the three vectors $\vec{v}_\perp(0)$, $\vec{v}'_\perp(0)$, and the momentum $\hat{p} \cdot \vec{v}_\perp(0)$ and $\vec{v}'_\perp(0)$ have the same magnitude and are $\pi/2$ out of phase.

Given these definitions, the time variation $\vec{v}_\perp(t)$ can now be written as

$$\vec{v}_\perp(t) = \vec{v}_\perp(0)e^{i\omega t} + \vec{v}'_\perp(0)e^{i(\omega t + \varepsilon\frac{\pi}{2})},$$

(A2.25)

where $\omega = H/\hbar$. It is important to keep in mind that despite the somewhat misleading vector notation, $\vec{v}_\perp(0)$ and $\vec{v}'_\perp(0)$ are operators. As vectors they and $\hat{p}$ only have one component as seen in Fig. (A2.1). The vector notation is useful, however, since it implies that

$$\vec{v}_\perp(0) \cdot [\hat{p} \times \vec{v}_\perp(0)] = [\vec{v}_\perp(0) \times \hat{p}] \cdot \vec{v}_\perp(0), \text{ or } \vec{v}_\perp(0) \cdot [\hat{p} \times \vec{v}_\perp(0)] + [\hat{p} \times \vec{v}_\perp(0)] \cdot \vec{v}_\perp(0) = 0,$$

which is the same as $\vec{v}_\perp(0) \cdot \vec{v}'_\perp(0) + \vec{v}'_\perp(0) \cdot \vec{v}_\perp(0) = 0$. The operators $\vec{v}_\perp(0)$ and $\vec{v}'_\perp(0)$ thus obey the anti-commutation relation and therefore do not commute.[28]

If the exponentials in Eq. (A2.25) are expanded, the real part is taken, $\vec{v}(t)$ may be written as

$$\vec{v}(t) = \vec{v}_\parallel(0) + \vec{v}_\perp(0)\cos\omega t - \vec{v}'_\perp(0)\varepsilon \sin\omega t.$$

(A2.26)

This equation may now be integrated, giving

$$\vec{x}(t) - \left[\vec{x}(0) - \varepsilon \frac{\vec{v}'_\perp(0)}{\omega}\right] = \vec{v}_\parallel(0)\,t + \frac{\vec{v}_\perp(0)}{\omega}\sin\omega t + \varepsilon \frac{\vec{v}'_\perp(0)}{\omega}\cos\omega t.$$

(A2.27)

Kobe calls the second term on the left hand side the "constant displacement". $\vec{v}_\parallel(0)\,t$ is corresponds to the displacement along the direction of the constant momentum $\hat{p}$. The amplitude of the zitterbewegung is $c/\omega$ and is a consequence of the magnitude of the eigenvalues of the velocity operators being $c$. $c/\omega$ is, of course, the wavelength divided by $2\pi$. But since the wavelength depends on the frame of reference, this is a good example of the non-localizability of the photon. The last two terms explicitly display the zitterbewegung.

## APPENDIX 3: Spontaneous Electroweak Symmetry Breaking

Spontaneous symmetry breaking is most easily explained by looking at *global* symmetry breaking, which means that gauge transformations are not space-time dependent. The simplest example is that of U(1). The general Lagrange density for a complex scalar field $\phi = (\phi_1 + i\phi_2)$ is

$$\mathcal{L} = (\partial_\mu \phi)(\partial^\mu \phi^*) - V(\phi\phi^*).$$

(A3.1)

For the potential $V$, one chooses a form originally proposed by Ginzburg and Landau before the BCS theory of superconductivity. This type of potential was intended to represent the Helmholtz free energy of a second order phase transition. The reason for choosing it here is that this form of potential works to give the desired result (and possibly tells us something about the nature of the vacuum) even though it was intended as a phenomenological description of the free energy density of a superconductor. In gauge theory it provides a type of self-interaction of the Higgs field. It is given by

$$V(\phi, \phi^*) = \mu^2 \phi^* \phi + \lambda (\phi^* \phi)^2.$$

(A3.2)

The self-interaction comes from the $\lambda$ term. The extrema of this function are found by taking the first and second derivative with respect to $|\phi|$ and setting the result equal to zero. Doing the algebra (and using the definition of $\phi$) results in

$$-\frac{\mu^2}{2\lambda} = \phi^* \phi = \phi_1^2 + \phi_2^2 =: a^2.$$

(A3.3)

$\phi\phi^* = |\phi|^2$, and $a^2$ is real for the choice $\lambda > 0$, $\mu^2 < 0$, which we make here. There is also the solution $\phi = \phi^* = 0$. Examining the second derivative tells us that this solution is a relative maximum and that the solution at $a^2 = -\mu^2/2\lambda$ is a relative minimum. In quantum field theory, $\phi$ becomes an operator whose minimum corresponds to the vacuum expectation value $|\langle 0|\phi|0\rangle|^2 = a^2$. A sketch of the potential given in Eq. (A3.2) is shown in Fig. (A3.1).

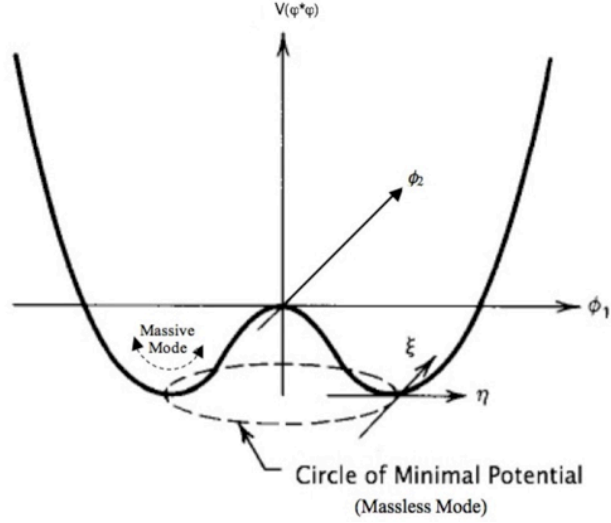

Figure (A3.1). A sketch of the Higgs potential with $\lambda > 0$ and $\mu^2 < 0$. Although the components of $\phi$ are drawn as coordinates, it should be remembered that $\phi$ is a field. The minima of the potential lie along the circle of minimal potential of radius $a$ that comprise a set of degenerate vacua related by a rotation about the axis corresponding to the magnitude of the potential. The potential along the circle, in the $\xi$ direction tangent to the circle, is constant. It therefore takes no energy to move along this path and motion along it corresponds to the massless mode, while motion in a plane containing the $V$-axis does take energy and corresponds to the massive mode.

Let us now transform to polar coordinates so that

$$\phi(x) = \rho(x)e^{i\theta(x)},$$

(A3.4)

where $x$ is the space-time coordinate. The vacuum is then $\langle 0|\phi|0\rangle = \langle 0|\rho|0\rangle = a$ and $\langle 0|\theta|0\rangle = 0$. The degenerate vacua are then connected by a U(1) symmetry transformation. Note that the U(1) phase symmetry is destroyed as a result of the vacuum being given by the choice of $\rho = a$ and some particular value of $\theta$; it is the specification of $\theta$ that breaks the symmetry. We will be interested in small oscillations around the vacuum state located at the circle of minimal potential. The quanta of these oscillations correspond to physically interesting particles. Because the minimum of the potential lies at a radial distance $a$ from the origin, the following transformation is made:

$$\phi(x) = \left(\rho'(x) + a\right)e^{i\theta(x)}.$$

(A3.5)

As a result the vacuum is now $\langle 0|\rho'|0\rangle = \langle 0|\theta|0\rangle = 0$. If this $\phi$ is then substituted into the Lagrangian given in Eqs. (A3.1) and (A3.2) there results, after a bit of algebra, a kinetic term and the potential term

$$V = \lambda\left(\rho'^4 + 4a\rho'^3 + 4a^2\rho'^2 - a^4\right).$$

(A3.6)

The quadratic term in $\rho'$ implies that $\rho'$ has a mass of $4\lambda a^2$. Spontaneous symmetry breaking has generated this mass. Notice that there is no similar term in $\theta^2$, implying that $\theta$ is a massless field. This can be thought of as being a consequence of there being no restoring force in the $\theta$-direction. $\phi_1$ and $\phi_2$ started out as two fields satisfying the Klein-Gordon equation. After symmetry breaking we have a massive field $\rho'$ and a massless field $\theta$. Such massless fields are know as Goldstone bosons.

The electroweak spontaneous symmetry breaking in the Standard Model has to do with the breaking of a *local* SU(2) gauge symmetry, which results in the elimination of the unwanted Goldstone bosons. This process is known as the Higgs mechanism. In this case, one can write the Higgs field as a single Higgs doublet in the unitary or "$u$" gauge as[29]

$$\phi = \frac{1}{\sqrt{2}}\begin{pmatrix} 0 \\ v + h(x) \end{pmatrix} \quad \phi_0 = \langle\phi\rangle = \frac{1}{\sqrt{2}}\begin{pmatrix} 0 \\ v \end{pmatrix},$$

(A3.7)

where $v = \sqrt{-\mu^2/\lambda}$ and $\phi_0$ is the minimum value of $\phi$. The small oscillations of the Higgs field $h(x)$ around the vacuum state located at the circle of minimal potential average to zero.

Under spontaneous symmetry breaking, the interaction Lagrangian for the first family of leptons ($e$, $\nu_e$, $u$, $d$) in the "$u$" gauge is

$$\mathcal{L}_{int} = \frac{G_e v}{\sqrt{2}}(\overline{e}_L e_R + \overline{e}_R e_L) + \frac{G_u v}{\sqrt{2}}(\overline{u}_L u_R + \overline{u}_R u_L) + \frac{G_d v}{\sqrt{2}}(\overline{d}_L d_R + \overline{d}_R d_L),$$

(A3.8)

where, for example, $e$ and $\overline{e}$ represent the appropriate Dirac spinors. The point is, that the electron and quarks have acquired the masses

$$m_i = \frac{G_i v}{\sqrt{2}}, \quad i = e, u, d.$$

(A3.9)

The masses of the fermions appear as parameters in the theory and must be put in by hand. The same is true for the mass of the Higgs boson, which can be found as follows:

The Lagrangian of a Lorentz invariant scalar field can be written as

$$\mathcal{L} = \frac{1}{2}(\partial_\mu \phi)(\partial^\mu \phi) - \frac{1}{2}m^2 \phi^2.$$

(A3.10)

If the Lagrangian for a scalar field with a symmetry breaking potential is simplified to

$$\mathcal{L}' = T - V = \frac{1}{2}(\partial_\mu \phi)^2 - \left(\frac{1}{2}\mu^2 \phi^2 + \frac{1}{4}\lambda \phi^4\right),$$

(A3.11)

and $\phi(x) = v + h(x)$ is substituted, one obtains

$$\mathcal{L}'' = \frac{1}{2}(\partial_\mu h)^2 - \lambda v^2 h^2 - \ldots .$$

(A3.12)

Comparing this to the Lagrangian in Eq. (A3.10) suggests that $m_h^2 = 2\lambda v^2$.

In the Standard Model, the Higgs mass $m_h$ is related to the values of the *W* boson and the top quark. The ATLAS AND CMS Collaborations[30] have experimentally determined the Higgs mass by using the decay channels $h \to ZZ \to$ (a combination of four electrons or muons whose total charge adds up to zero), and $h \to \gamma\gamma$. The result was $m_h$ = 125.09 GeV.